\documentclass[]{spie}  
\usepackage{amsmath,amsfonts,amssymb}
\usepackage{graphicx}
\usepackage{setspace}
\usepackage{tocloft}
\usepackage{lineno}
\usepackage{url}
\usepackage[colorlinks=true, allcolors=blue]{hyperref}

\title{Demonstrating sub-electron noise performance in Single electron Sensitive Readout (SiSeRO) devices
}

\author[a]{Tanmoy Chattopadhyay}
\author[a]{Sven Herrmann}
\author[a]{Peter Orel}
\author[b]{Kevan Donlon}
\author[a,c,d]{Steven W. Allen}
\author[e]{Marshall W. Bautz}
\author[a]{Brianna Cantrall}
\author[b]{Michael Cooper}
\author[e]{Beverly LaMarr}
\author[b]{Chris Leitz}
\author[e]{Eric Miller}
\author[a]{R. Glenn Morris}
\author[a,d]{Abigail Y. Pan}
\author[e]{Gregory Prigozhin}
\author[b]{Ilya Prigozhin}
\author[a,d]{Haley R. Stueber}
\author[a]{Daniel R. Wilkins}

\affil[a]{Kavli Institute of Astrophysics and Cosmology, Stanford University, 452 Lomita Mall, Stanford, CA 94305, USA}
\affil[b]{MIT Lincoln Laboratory, Lexington, MA, USA}
\affil[c]{SLAC National Accelerator Laboratory, 2575 Sand Hill Road, Menlo Park, CA 94025, USA}
\affil[d]{Department of Physics, Stanford University, 382 Via Pueblo Mall, Stanford CA 94305, USA}
\affil[e]{Kavli Institute for Astrophysics and Space Research, Massachusetts Institute of Technology, Cambridge, MA, USA}

\authorinfo{Further author information: (Send correspondence to Tanmoy Chattopadhyay)\\T. Chattopadhyay: E-mail: tanmoyc@stanford.edu}

\begin{document} 
\maketitle

\begin{abstract}
Single electron Sensitive Read Out (SiSeRO) is a novel on-chip charge detection technology that can, in principle, provide significantly greater responsivity and improved noise performance than traditional charge coupled device (CCD) readout circuitry. The SiSeRO, developed by MIT Lincoln Laboratory, uses a p-MOSFET transistor with a depleted back-gate region under the transistor channel; as charge is transferred into the back gate region, the transistor current is modulated. With our first generation SiSeRO devices, we previously achieved a responsivity of around 800 pA per electron, an equivalent noise charge (ENC) of 4.5 electrons root mean square (RMS), and a full width at half maximum (FWHM) spectral resolution of 130 eV at 5.9 keV, at a readout speed of 625 Kpixel/s and for a detector temperature of 250 K. Importantly, since the charge signal remains unaffected by the SiSeRO readout process, we have also been able to implement Repetitive Non-Destructive Readout (RNDR), achieving an improved ENC performance. In this paper, we demonstrate sub-electron noise sensitivity with these devices, utilizing an enhanced test setup optimized for RNDR measurements, with excellent temperature control, improved readout circuitry, and advanced digital filtering techniques. 
We are currently fabricating new SiSeRO detectors with more sensitive and RNDR-optimized amplifier designs, which will help mature the SiSeRO technology in the future and eventually lead to the pathway to develop active pixel sensor (APS) arrays using sensitive SiSeRO amplifiers on each pixel. Active pixel devices with sub-electron sensitivity and fast readout present an exciting option for next generation, large area astronomical X-ray telescopes requiring fast, low-noise megapixel imagers.
  
\end{abstract}





\section{INTRODUCTION}
\label{sec:intro}  

X-ray Charge Coupled Devices (CCDs)  \cite{Lesser15_ccd,gruner02_ccd,ccd_janesick01} have been the workhorses of X-ray astronomical instrumentation for more than three decades. 
With recent technological progress and the development of sensitive readout electronics, current generation X-ray CCDs provide significantly improved readout speeds ($\sim$2 MHz) and noise performance \cite{bautz18,bautz19,bautz20,chattopadhyay22_ccd,Hermannetal2022,herrmann20_mcrc,Bautzetal2022,Oreletal2022}. 
However, to enable next generation flagship missions in X-ray astronomy (e.g. Lynx\footnote{\url{https://www.lynxobservatory.com/}}), future X-ray detectors will require much faster readout rates 
and very low read noise, ideally in the sub-electron regime. With such low noise, the full discovery potential at soft X-ray energies (E~$<0.5$ keV) can be achieved, together with unprecedented gain calibration \cite{rodrigues21}. Precise gain calibration at low X-ray energies is also helpful in characterizing the low energy response of instruments accurately. 

The Single electron Sensitive Read Out (SiSeRO) technology developed by MIT Lincoln Laboratory (MIT-LL) is a novel charge detection approach for X-ray CCDs \cite{chattopadhyay22_sisero}. SiSeROs replace the doped implant at the sense node of the CCD output stage with an internal gate under a p-type metal-oxide-semiconductor field-effect transistor (p-MOSFET). The internal gate minimizes the parasitic capacitance on the sense node, providing higher gain and improved low noise performance. Perhaps the most interesting feature of SiSeRO devices is that the signal charge can be transferred in and out of the internal gate without destroying the charge $-$ a technique popularly known as repetitive non-destructive readout (RNDR). Utilizing RNDR, read noise can be reduced significantly. The RNDR technique has previously been demonstrated and shown to provide sub-electron noise yield in DEPFET devices \cite{wolfel06,treberspurg22_rndrdepfet}, Skipper CCDs \cite{tiffenberg17,rodrigues21,Lapi2024_skipperMAS} and in a prototype Complementary Metal-Oxide-Semiconductor (CMOS) imaging sensor (CIS) \cite{Stefanov2020_skippercmos,Lapi2024_skipperCMOS}. 

In Chattopadhyay et al. 2024\cite{chattopadhyay24_rnrdr}, we presented the first demonstration of the RNDR technique in a SiSeRO prototype device. That study implemented nine RNDR cycles, where each cycle was $\sim$1.0 $\mu$s CDS (correlated double sampling) long, achieving an equivalent noise charge (ENC) of 2.06 $\mathrm{e}^{-}_{\mathrm{RMS}}$ at the end of the ninth cycle (a factor of 3 improvement in noise). In this paper, we report RNDR measurements for a larger number of cycles. The new tests were performed using an improved experimental setup, which provides enhanced cooling of the detectors. With around 1 $\mu$s CDS per cycle, and 57 RNDR cycles, we achieved a readout noise of 0.51 $\mathrm{e}^{-}_{\mathrm{RMS}}$.      
While the current SiSeRO prototypes are in the CCD format, the technology also has the potential to be incorporated into active pixel sensor (APS) arrays, in which each pixel includes a SiSeRO. This should enable extremely low noise performance at high readout speeds. A prototype SiSeRO APS device is currently being fabricated at MIT-LL and will be tested early next year. 
In parallel, in order to test new sensitive SiSeRO amplifier designs, we are fabricating new SiSeRO based CCD devices with 16 different SiSeRO variants at the output. Along with that, an RNDR optimized SiSeRO CCD variant is also under fabrication offering sixteen SiSeRO amplifiers in series at the output stage. Such devices should immediately provide a factor of four improvement in noise performance at the same readout rate. 

In the next sections, we give a brief description of the SiSeRO devices and the RNDR technique, the experimental setup, and the new RNDR results. We summarize the work in section \ref{sec:summary} and outline our future plans.


\section{An overview of the SiSeRO devices and the RNDR technique}
\begin{figure}[t!]
    \centering
   \includegraphics[width=.48\linewidth]{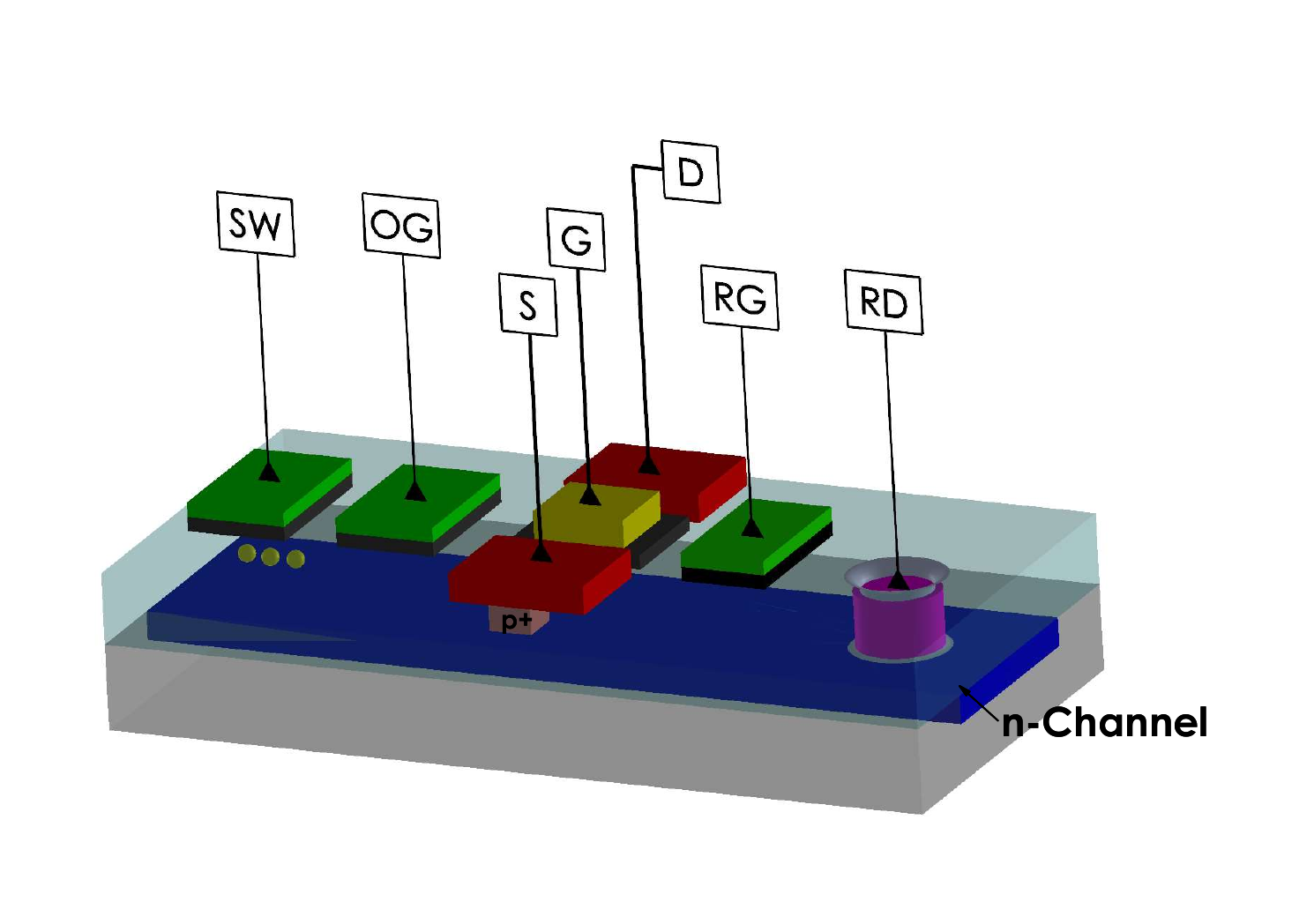}
   \includegraphics[width=.48\linewidth]{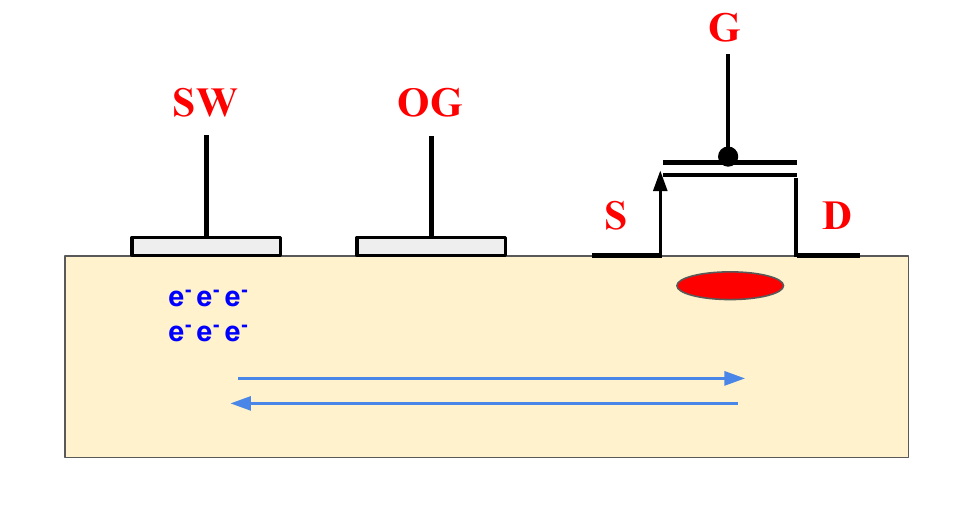}
    \caption{Left: Schematic of a SiSeRO output stage, utilizing a p-MOSFET transistor (S: Source, D: Drain, G: Gate) with an internal gate beneath the p-MOSFET. The internal gate is emptied using the reset transistor (RG) and reset drain (RD). Charge transferred across the output gate (OG) from the last serial gate (summing well or SW) to the internal gate modulates the drain current of the p-MOSFET. Right: A simple demonstration of RNDR technique. The signal charge is moved between the internal gate and SW for repeated measurements.
    }
    \label{fig:sisero}
\end{figure}
The current SiSeRO detectors, fabricated at MIT-LL, consist of a CCD pixel array with a SiSeRO amplifier  (p-MOSFET) at the output stage. Figure \ref{fig:sisero} (left) shows a simplified 3D schematic of the SiSeRO output stage. When a charge packet is transferred from summing well (SW, the last serial register gate) to the internal gate via OG (output gate), it modulates the transistor drain current, which is then readout by the readout electronics \cite{chattopadhyay22_sisero,Chattopadhyayetal2022}. There is a trough on the internal gate that helps confine the charge packet under the MOSFET to maximize the change in the transitor current. 
An advantage of SiSeROs over conventional voltage follower-based CCD output stages is that the internal gate minimizes parasitic capacitance on the sense node, resulting in a high conversion gain and minimized noise. 

Our first generation SiSeRO test devices have an active area of $\sim$4 mm $\times$ 4 mm with a 512 $\times$ 512 array of 8 $\mu$m pixels. They come in two primary variants $-$ surface channel and buried channel p-MOSFET transistors, along with sub-variants that adjust the location and size of the trough. Unsurprisingly, the surface channel MOSFET variants (CCID-85 family) were found to have relatively large readout noise \cite{chattopadhyay22_sisero}. 
For variants with a buried-channel SiSeRO (CCID-93 family), we initially achieved an ENC of 6 $\mathrm{e}^{-}_{\mathrm{RMS}}$ at 625 KHz readout speed (1 $\mu$s CDS). This was further improved to 4.5 $\mathrm{e}^{-}_{\mathrm{RMS}}$ by applying digital filtering techniques to minimize the 1/f noise \cite{chattopadhyay23_sisero}. We measured a gain or responsivity of 800 pico-ampere (pA) per electron for this device. The spectral resolution at the energy of the 5.9 keV Mn K$_\alpha$ line was measured to be $\sim$ 132 eV (FWHM).  

As noted above, one of the most interesting features of SiSeRO devices is that the charge packet in the internal gate remains unaffected during the readout process, which offers the possibility of repetitive measurements of the same charge signal by moving the charge in and out of the internal gate multiple times. By averaging over a large number of multiple independent measurements of the same charge, the final noise can be reduced significantly, potentially into the sub-electron regime. 
In Fig. \ref{fig:sisero} (right), we show a simplified schematic of how RNDR is implemented in our setup. In the absence of dedicated transfer gates in the current prototype devices, we employed clocking to the output gate (OG) and summing well (SW) to pull the charge back from the internal gate. 
In Chattopadhyay et al. 2024\cite{chattopadhyay24_rnrdr}, we reported RNDR measurements for nine repetitive readout cycles with 1 $\mu$s CDS per cycle integration. 
With nine RNDR cycles on a CCID-93 device, we obtained an ENC performance of 2.06 $\mathrm{e^{-}_{RMS}}$ at the end of the ninth cycle \cite{chattopadhyay24_rnrdr}. In this study, we perform a larger number of cycles (57) on a different SiSeRO detector from the same CCID-93 family. 


\section{Experimental setup} \label{sec:setup}
\begin{figure}[t]
    \centering
    \includegraphics[width=0.62\linewidth]{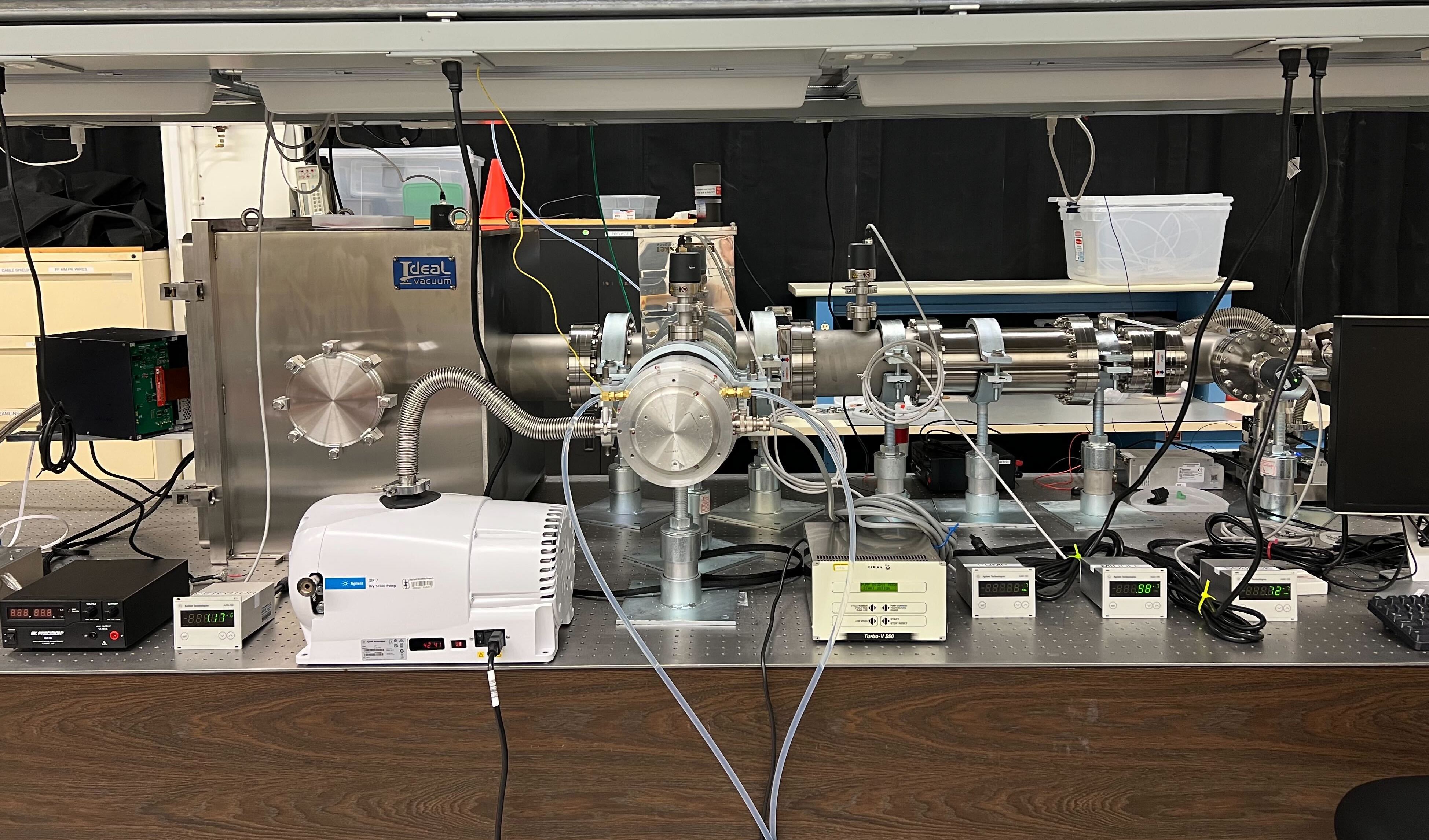}
    \includegraphics[width=0.37
    \linewidth]{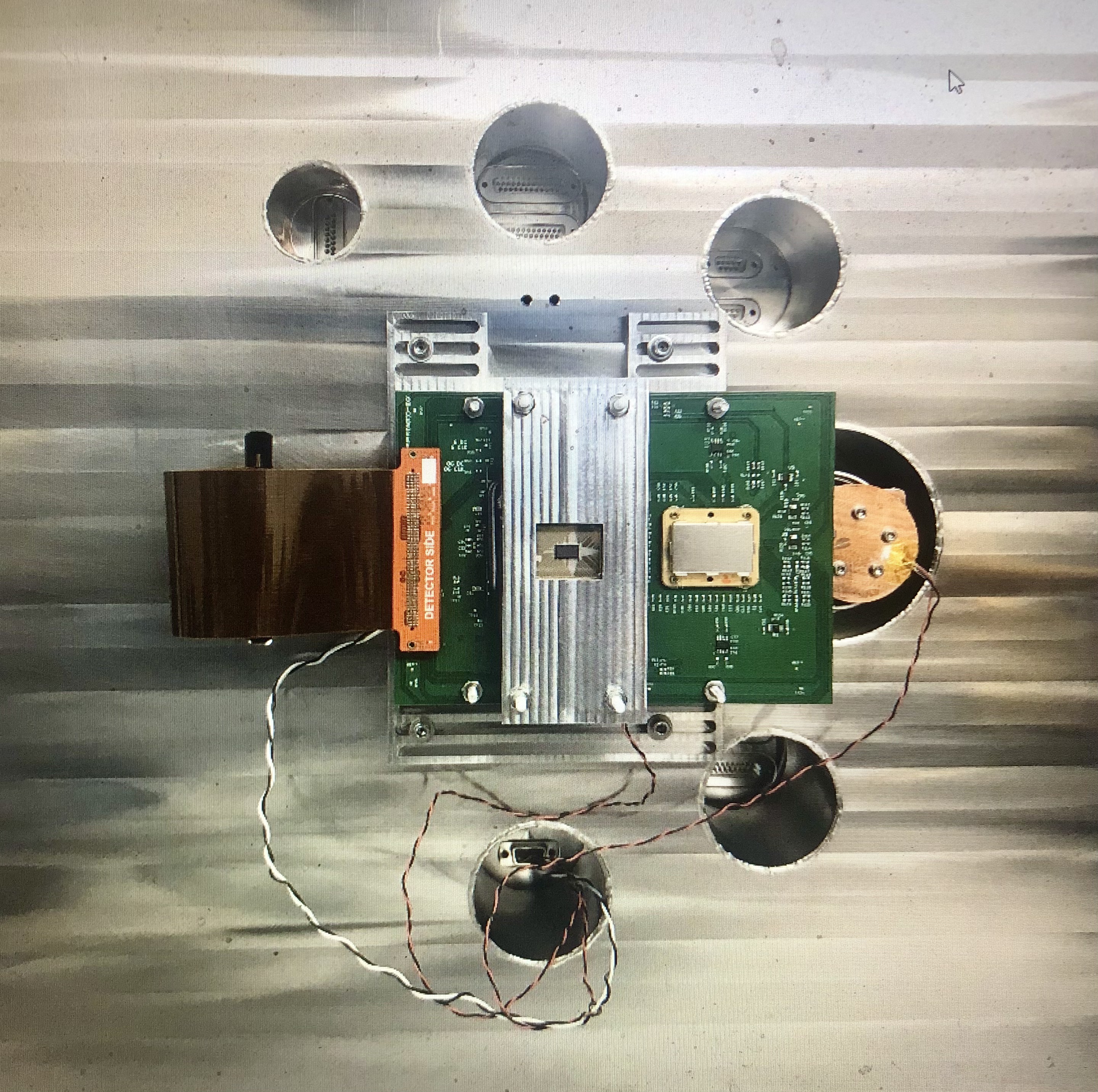}
    \caption{Left: The 2.5 meter XOC X-ray Beamline\cite{stueber2024} used for RNDR experimentation. Right: A CCID-93 SiSeRO device mounted inside the detector chamber on the chamber door.}
    \label{fig:setup}
\end{figure}

Our most recent SiSeRO measurements were carried out in the newly-commissioned 2.5 meter XOC X-ray Beamline at Stanford \cite{stueber2024} (hereafter beamline) shown in Fig. \ref{fig:setup}. 
The beamline can be used for the testing and characterization of a broad range of X-ray detectors of different types and sizes. The detector under test is mounted directly on the inside of the detector chamber door (Fig. \ref{fig:setup} right). The beamline employs a cryo-cooler to cool down the detector to -100$^\circ$C.  
A Proportional-Integral-Derivative (PID) algorithm controls the detector temperature with better than 0.2$^\circ$C accuracy. An $^{55}$Fe radioactive source is mounted within the detector chamber, and an X-ray fluorescence (XRF) source module is mounted at the other end of the beamline. The XRF source can provide mono-energetic lines spanning the 0.3$-$10 keV range at suitable flux levels. Details of the beamline mechanical design, detector cooling, and source module can be found in Stueber et al. 2024\cite{stueber2024}.   

As mentioned earlier, for this experiment, we used a CCID-93 detector with a buried-channel SiSeRO at the output stage. 
The readout electronics \cite{chattopadhyay22_sisero,chattopadhyay24_rnrdr} consist of a custom preamplifier board (the detector and the circuit board are shown in Fig. \ref{fig:setup} right) and a commercial Archon controller \cite{archon14} (the black box in the left picture). The preamplifier uses a drain current readout, where an I2V amplifier first converts and amplifies the SiSeRO output and a differential driver (at the second stage) converts the output to a fully differential signal.
The Archon, procured from Semiconductor Technology Associates, Inc (STA\footnote{\url{http://www.sta-inc.net/archon/}}), provides the required bias and clock signals, and digitization of the output signal. A custom flex cable \cite{porelMCRCspie2024} is used to connect the detector board to the controller which is kept outside the detector chamber. We have also developed an 8-channel Application-Specific Integrated Circuit (ASIC) based readout electronics (known as Multi-Channel Readout Chip or MCRC ASIC\cite{herrmann20_mcrc,Oreletal2022}) for the SiSeROs. The results presented in this paper were obtained from the discrete electronics readout module.  

\section{Results}
\label{sec:results}
\begin{figure}
 \centering
   \hspace{-0.3cm}
    \includegraphics[width=.8\linewidth]{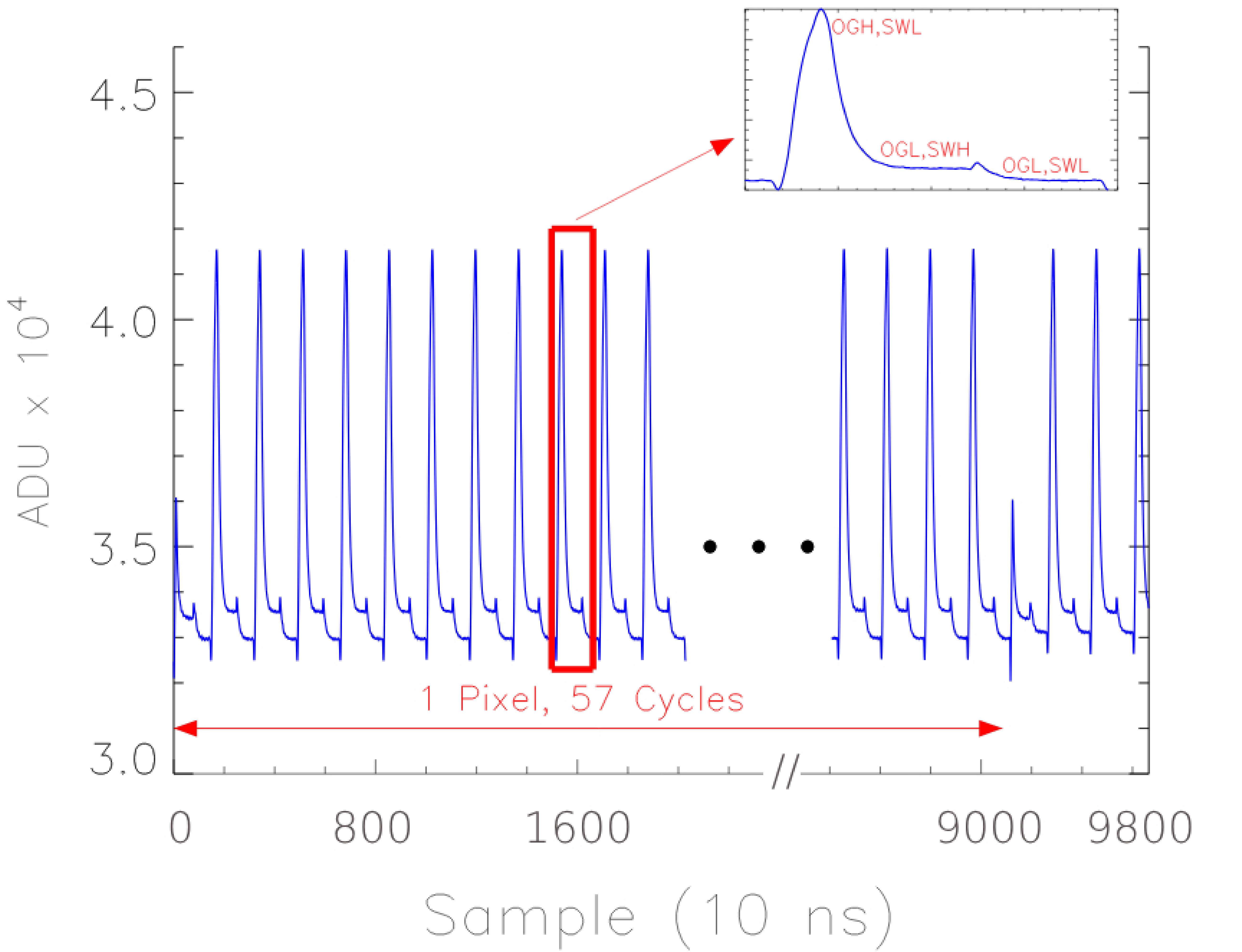}
    \caption{SiSeRO digital waveform for one pixel over 57 repetitive readout cycles showing the baseline and signal measurements for each cycle. The zoom-in view shows the OG clock, baseline and signal for a single RNDR cycle.}
    \label{fig:rndr_waveform}
\end{figure}
Figure \ref{fig:rndr_waveform} shows waveform of the SiSeRO for one pixel over 57 RNDR cycles. As noted above, for the RNDR measurements, we clock OG and SW gates (OGhigh 5.5 V, OGlow 1.0 V; SWhigh 3.0 V, SWlow -1.5 V). The individual baselines are introduced when OG settles down to OGLow (see the insert plot in Fig. \ref{fig:rndr_waveform}). Note that the OG clock pulses in the waveform result from some capacitive coupling between OG and the MOSFET. The corresponding individual signal regions are introduced when the charge packet is moved back to the internal gate by changing the SW potential to SWLow (as shown in the insert plot). 
After the final cycle, the charge packet is drained to RD using the reset gate (RG) clock.

For our tests, we used a titanium (Ti) target in the XRF source, providing K$_\alpha$ and K$_\beta$ lines at 4.5 and 4.9 keV respectively. 
\begin{figure}[t]
    \centering
 \includegraphics[width=.48\linewidth]{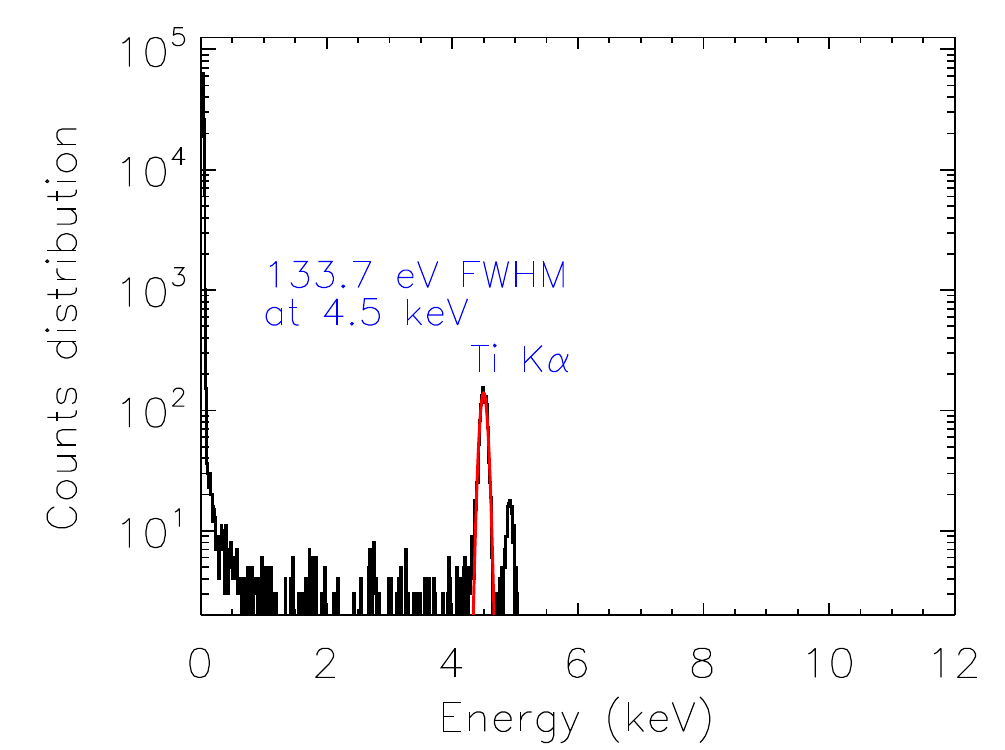}
 \includegraphics[width=.48\linewidth]{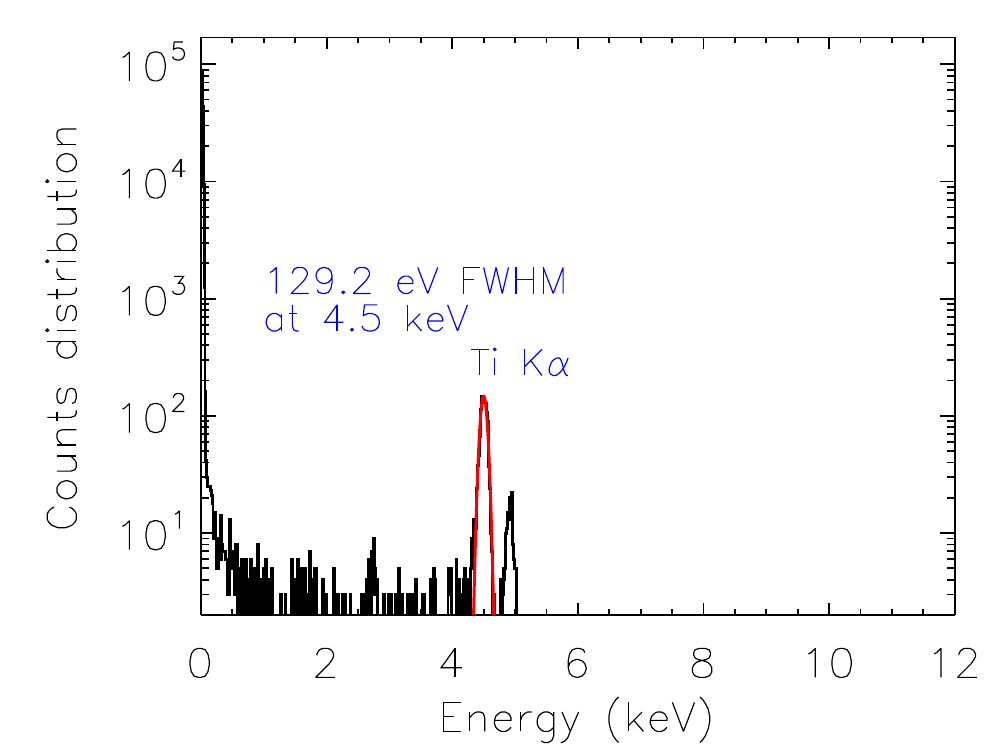}
    \caption{Spectrum obtained at -100$^\circ$C at the beginning (1$\mathrm{^{st}}$ RNDR cycle) (left) and at the end of the 57$\mathrm{^{th}}$ RNDR cycle (right) showing the Ti K$_\alpha$ (4.5 keV) and K$_\beta$ (4.9 keV) lines. The spectra were generated using the single-pixel (grade 0) events. The FWHM energy resolutions at 4.5 keV are 134 eV and 129 eV and the corresponding read noises 3.8 and 0.51 $\mathrm{e}^{-}_{\mathrm{RMS}}$, respectively.}
    \label{fig:spectra}
\end{figure}
For each RNDR cycle, we generate single-pixel (grade 0) spectra of the K$_\alpha$ and K$_\beta$ lines and fit the 4.5 keV line with a Gaussian model to calculate the centroid and FWHM of the line. We also calculate the read noise for each cycle by measuring the width of distribution of charge (in ADU or analog to digital unit) in the overscan region \cite{chattopadhyay24_rnrdr}. Because of minimal thermal leakage contamination, the noise measured from the overscan region supposedly represents the noise due to the output stage and the following readout circuitry. The measured gains are then used to normalize and measure the read noise in number of electrons.  

\begin{figure}[h]
    \centering
   \includegraphics[width=.49\linewidth]{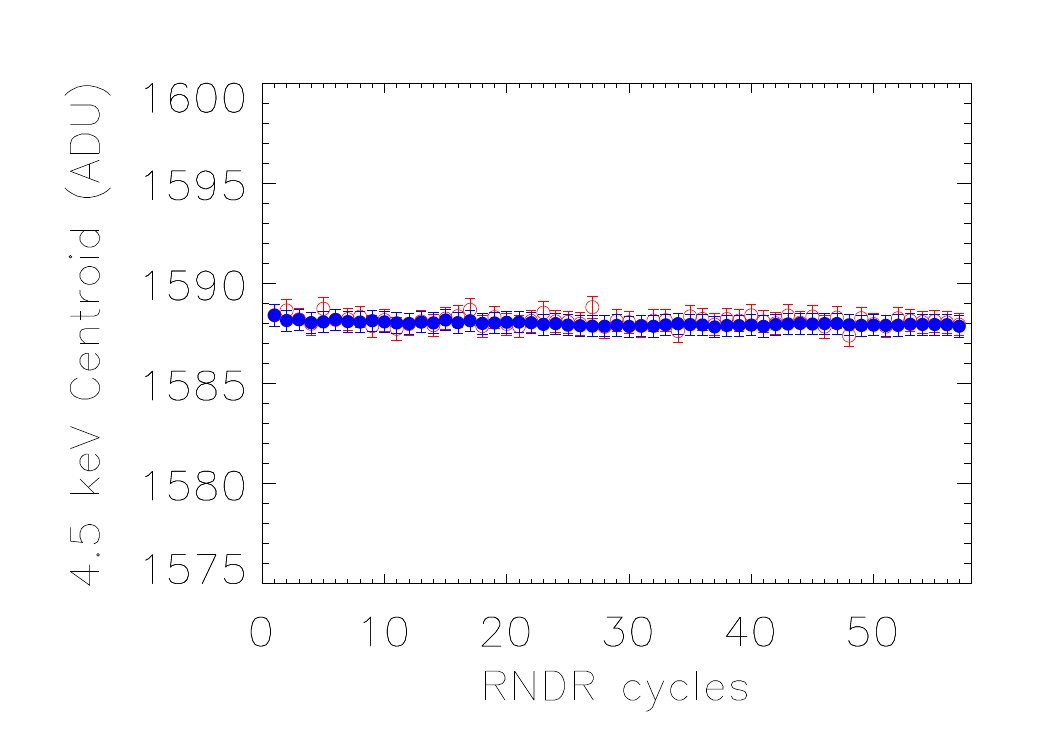}
   \includegraphics[width=.49\linewidth]{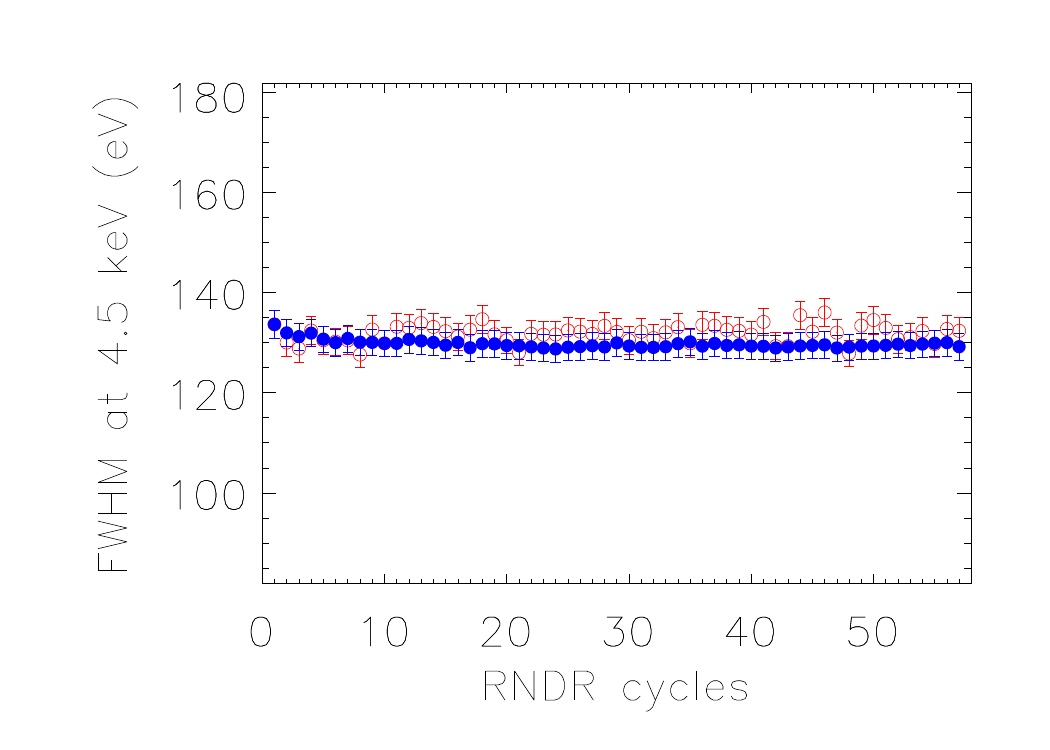}
    \caption{Left: Measured centroid in analog to digital unit (ADU) for 4.5 keV as a function of RNDR cycles (representing gain). Both the individual (blue circles) and averaged gains (red circles) remain constant, implying no charge loss during the repetitive transfer of charge. Right: Measured FWHM at 4.5 keV as a function of RNDR cycles. Both the individual and the averaged measurements lie well within the $\sim$1$\sigma$ error bar of each other. }
    \label{fig:rndr_results1}
\end{figure}
As an example, the spectra along with the Gaussian fits, obtained for the 1$\mathrm{^{st}}$ and 57$\mathrm{^{th}}$ RNDR cycles are shown in the left and right panel of Fig. \ref{fig:spectra} respectively.
The 4.5 keV line centroids (representation of gain of the system), measured in this way for all the individual RNDR cycles, are shown by the red circles in the left plot of Fig. \ref{fig:rndr_results1}. The blue circles show the averaged measurements of the same. The centroid values are found to be well within the error bar of each other which implies that there is no charge loss during the fifty seven repetitive transfers of charge. Using the fitted centroid values, we also calculate the responsivity of the device to be around 830 pA / electron. 

The measured FWHM values of the 4.5 keV line are shown as a function of the RNDR cycles in the right plot of Fig. \ref{fig:rndr_results1}. The individual measurements (blue circles) all lie within the $\sim$1$\sigma$ error bar of each other at aroud 134 eV. The averaged measurements, shown by the blue cirles in the same plot, show a slight improvement in the spectral resolution at around 129 eV. 

\begin{figure}[h]
    \centering
   \includegraphics[width=.8\linewidth]{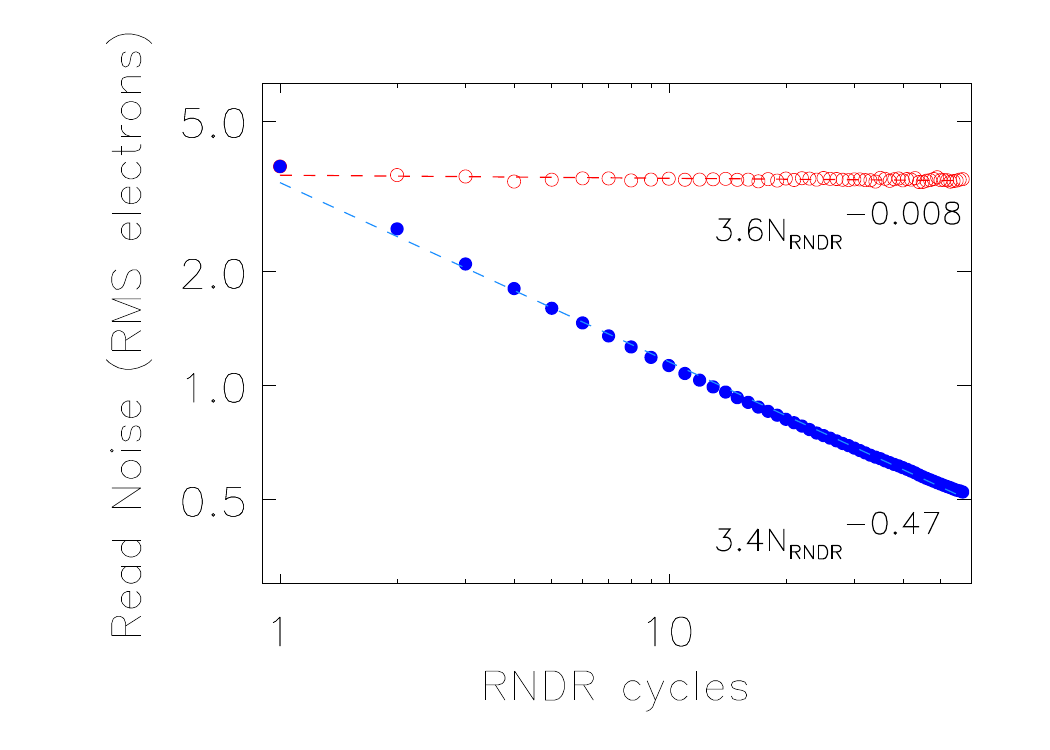}
    \caption{Read noise from the RNDR analysis. The red circles show the individual read noise measurements from the 57 individual measurements while the blue filled circles are obtained after averaging over the measurements. The measurements follow the expected 1/$\mathrm{\sqrt{N}}$ trend, shown by the dashed blue line. The read noise for the first cycle is measured to be 3.8 $\mathrm{e}^{-}_{\mathrm{RMS}}$, while at the end of fifty seventh cycle, the read noise has improved to 0.51 $\mathrm{e}^{-}_{\mathrm{RMS}}$. }
    \label{fig:rndr_results2}
\end{figure}

The read noise measurements are demonstrated in Fig. \ref{fig:rndr_results2}, where we show the individual and averaged measurements in red and blue circles respectively. As seen in the figure, the individual cycles yield similar noise measurements around 3.8 $\mathrm{e}^{-}_{\mathrm{RMS}}$. This is cross-verified by fitting the values with a power law, which results in a flat line (shown by the red dashed line). This further suggests that the transfer of charge in and out of the internal gate is efficient.       
The averaged measurements show that the noise follows a nearly 1/$\mathrm{\sqrt N}$ trend, as shown by the fitted power law line (blue dashed line). A slight deviation from the 1/$\mathrm{\sqrt N}$ trend is expected due to the presence of 1/f noise in these detectors \cite{sofoharo2023_nsisero,moroni2012_skipper}. Starting with an ENC of 3.8 $\mathrm{e}^{-}_{\mathrm{RMS}}$, we achieved an ENC of 0.51 $\mathrm{e}^{-}_{\mathrm{RMS}}$ at the end of the 57$\mathrm{^{th}}$ RNDR cycle.
     

\section{Summary and future plans}\label{sec:summary}

Single electron Sensitive Read Out (SiSeRO) is a novel on-chip charge detection technology that can provide improved responsivity and noise performance than traditional readout circuitry. Developed by MIT-LL, the current SiSeRO prototype devices are in the CCD format with each output stage having a SiSeRO amplifier. In this paper, we demonstrated sub-electron noise performance for a CCID-93 SiSeRO prototype by implementing 57 RNDR cycles. The noise evolution across the cycles show a nearly $\mathrm{1/\sqrt{N}}$ trend, with a final noise of 0.51 $\mathrm{e^{-}_{RMS}}$ at an operating temperature -100$^\circ$C and an impressive readout rate of 1 $\mu$s of CDS integration per cycle. 

In order to mature the SiSeRO technology, we are currently fabricating three novel SiSeRO detectors at MIT-LL based on more optimized and improved SiSeRO amplifier designs \cite{Donlonpie2024}. The first detector variant will have 16 parallel SiSeROs at the output, with variations in the geometry of the internal gate, location of the trough, and isolation of the MOSFET from the surrounding gates. This will allow us to optimize the technology further for better noise and gain performance. Another SiSeRO variant, aimed towards optimizing the RNDR method, will have 16 SiSeRO amplifiers in series at the output, with dedicated transfer gates between each pair of the SiSeRO amplifiers. This will allow 16 measurements of the same signal charge, resulting in a reduction of noise by a factor of four while maintaining the same frame rate. The noise can be reduced further by introducing RNDR sampling between each pair of the SiSeRO amplifiers. Both these detector types are expected to be fabricated by the end of the current year. We plan to utilize the MCRC ASIC to readout these detectors.   

Our long-term goal is to build a SiSeRO active pixel sensor (APS) following the concept used in RNDR optimized DEPFET detectors \cite{wolfel06}. 
The first prototype SiSeRO matrix is currently under fabrication and will be available for testing soon. Based upon the test results, we later plan to develop a larger matrix array based on this concept. Such a detector will provide a pathway to develop an extremely low noise, fast readout mega-pixel spectro-imager for future X-ray astronomical missions.          

\acknowledgments
This work has been supported by NASA grants APRA 80NSSC19K0499 ``Development
of Integrated Readout Electronics for Next Generation X-ray CCDs” and SAT
80NSSC20K0401 ``Toward Fast, Low-Noise, Radiation-Tolerant X-ray Imaging Arrays for
Lynx: Raising Technology Readiness Further.”






\end{document}